\documentclass[10pt,journal,compsoc]{IEEEtran}

\usepackage[sort&compress,numbers]{natbib}
\usepackage{amsmath,amssymb,amsfonts}
\usepackage{graphicx}
\usepackage{subcaption}
\usepackage{textcomp}
\usepackage{xcolor}
\usepackage{balance}
\usepackage{comment}
\usepackage{multirow}
\usepackage{siunitx}    %for \num
\usepackage{soul}
\usepackage{textcomp}
\usepackage{tabularx}
\usepackage{algorithm}
\usepackage{algpseudocode}
\usepackage{xspace}
\usepackage{natbib}
\usepackage{listings}
\usepackage{styles/c_custom_style}
\usepackage{styles/nasm_custom_style}
\newcommand{\shortnm}{{\bf IterPro}\xspace}

\newcommand{\armor}{{\bf Builder}\xspace}
\newcommand{\safeguard}{{\bf Runtime}\xspace}

\begin{document}

\title{Near-zero Downtime Recovery from Transient-error-induced Crashes}
%
%
% author names and IEEE memberships
% note positions of commas and nonbreaking spaces ( ~ ) LaTeX will not break
% a structure at a ~ so this keeps an author's name from being broken across
% two lines.
% use \thanks{} to gain access to the first footnote area
% a separate \thanks must be used for each paragraph as LaTeX2e's \thanks
% was not built to handle multiple paragraphs
%

\author{Chao Chen,
        Greg Eisenhauer,
        and~Santosh Pande% <-this % stops a space
\IEEEcompsocitemizethanks{
    \IEEEcompsocthanksitem This work was done when
    Chao Chen was a PhD student in the School of Computer 
    Science, Georgia Institute of Technology, 
    Atlanta, GA, 30332.\protect\\
    E-mail: chao.chen@gatech.edu
    
    \IEEEcompsocthanksitem Greg Eisenhauer and 
    Santosh Pande are with the School of Computer 
    Science, Georgia Institute of Technology, 
    Atlanta, GA, 30332.\protect\\
    E-mail: {eisen, santosh.pande}@cc.gatech.edu
  }% <-this % stops a space
% \thanks{Manuscript received April 19, 2005; revised August 26, 2015.}
}

% The paper headers
\markboth{Journal of IEEE Transactions on Parallel and Distributed Systems}%
{Shell \MakeLowercase{\textit{Chao Chen, Greg Eisenhauer and Santosh Pande}}: 
Near-zero Downtime Recovery from Transient-Error-Induced Crashes}
% The only time the second header will appear is for the odd numbered pages
% after the title page when using the twoside option.
% 
% *** Note that you probably will NOT want to include the author's ***
% *** name in the headers of peer review papers.                   ***
% You can use \ifCLASSOPTIONpeerreview for conditional compilation here if
% you desire.

% If you want to put a publisher's ID mark on the page you can do it like
% this:
%\IEEEpubid{0000--0000/00\$00.00~\copyright~2015 IEEE}
% Remember, if you use this you must call \IEEEpubidadjcol in the second
% column for its text to clear the IEEEpubid mark.

% use for special paper notices
%\IEEEspecialpapernotice{(Invited Paper)}

\IEEEtitleabstractindextext{
\begin{abstract}

Due to the system scaling, {\it transient errors} caused by external 
noises, e.g., heat fluxes and particle strikes, have become a growing 
concern for the current and upcoming extreme-scale high-performance-computing 
(HPC) systems. Applications running on these systems are expected to 
experience transient errors more frequently than ever before, which 
will either lead them to generate incorrect outputs or cause them to 
crash. However, since such errors are still quite rare as compared to 
no-fault cases, desirable solutions call for low/no-overhead systems 
that do not compromise the performance under no-fault conditions and 
also allow very fast fault recovery to minimize downtime. 
In this paper, we present \shortnm, a light-weight compiler-assisted 
resilience technique to quickly and accurately recover processes from 
transient-error-induced crashes. During the compilation of 
applications, \shortnm  constructs a set of recovery kernels 
for crash-prone instructions. These recovery kernels are 
executed to repair the corrupted process states on-the-fly upon 
occurrences of errors, enabling applications to continue their 
executions instead of being terminated. When constructing recovery 
kernels, \shortnm exploits side effects introduced by induction variable based code optimization 
techniques based on loop unrolling and strength reduction to improve 
its recovery capability. To this end, two new code transformation passes 
are introduced to expose the side effects for resilience purposes. 
We evaluated \shortnm with $4$ scientific workloads as well as the NPB benchmarks 
suite. During their normal execution, \shortnm incurs almost {\bf zero} runtime 
overhead and a small, fixed \textbf{27MB} memory overhead. Meanwhile, \shortnm can recover 
on an average $83.55\%$ of crash-causing errors within dozens of milliseconds with 
negligible downtime. 
% We 
% also evaluated \shortnm with parallel jobs running on $3072$ cores and showed that 
% \shortnm can successfully mask the impact of crash-causing errors by providing 
% almost uninterrupted execution. Finally, We present our preliminary evaluation 
% result for BLAS, which shows that \shortnm is capable of recovering failures in 
% libraries with a very high coverage rate of $83\%$ and negligible overheads. 
With such an effective recovery mechanism, \shortnm could tremendously mitigate 
the overheads and resource requirements of the resilience subsystem in future 
extreme-scale systems.

\end{abstract}

% Note that keywords are not normally used for peerreview papers.
\begin{IEEEkeywords}
Resiliency, Transient Fault, Soft Error, Fault Tolerance, Exa-scale Computing, Failure, Crash, Segment fault, Compiler
\end{IEEEkeywords}

}

% make the title area
\maketitle
\IEEEdisplaynontitleabstractindextext

% For peer review papers, you can put extra information on the cover
% page as needed:
% \ifCLASSOPTIONpeerreview
% \begin{center} \bfseries EDICS Category: 3-BBND \end{center}
% \fi
%
% For peerreview papers, this IEEEtran command inserts a page break and
% creates the second title. It will be ignored for other modes.
\IEEEpeerreviewmaketitle

\IEEEraisesectionheading{\section{Introduction}}
Reliability is a fundamental feature expected from extreme-scale high 
performance computing (HPC) systems, where a chance of failure of a system comprised of millions of cores
and other components running long running codes under extreme conditions of energy consumption
becomes significant. Moreover, as new computing architectures 
continue to boost system performance and energy efficiency with higher 
circuit density, shrinking transistor size and near-threshold voltage 
(NTV) operations, concern is growing in the HPC community about undesirable 
side-effects of these manufacturing trends, specifically in terms of increase in the transient errors 
caused by external noises, such as heat fluxes and high energy 
particles~\citep{DBMA2011,CGGK2014,HuLu2016}. Unfortunately, there 
is a lack of cost-efficient mechanisms to mask these errors at the hardware 
level~\citep{Hero2013,SDBF2015}, therefore applications running 
on these systems are expected to experience transient errors 
more frequently than ever before, and efficient application-level 
resilience techniques are required for future scientific 
applications~\citep{Hero2013,DRPG2014,MBCC2014,KFBF2016}.

In general, transient errors can result in two types of issues while executing
scientific applications. They could either lead applications to 
generate incorrect outputs (Silent Data Corruptions or SDCs) or 
cause them to crash (referred as soft failures in the rest of 
the paper)~\cite{LiVY2012,CGBM2014,FGDP2017,CSOG2017}.
While there has been significant amount of prior work on detecting 
and correcting SDCs~\citep{Chen2011,SDFC2016,CGMS2018}, less research 
effort has been spent on lightweight recovery of soft failures, perhaps 
because the community takes it for granted that the standard Checkpoint/Restart 
(C/R) methods can provide adequate recovery. Unfortunately, while the 
C/R technique is effective for recovery from these soft failures, it suffers 
from  {\it extreme costs} in terms of lost opportunities (batch job slots), lost 
computation (everything since the last checkpoint) and I/O overheads 
(repeatedly writing checkpoint files) and a significant slowdown under normal 
(no fault) execution of the applications. These costs are particularly 
significant for massively parallel jobs~\cite{DBMA2011,EKFF2012} in 
the HPC environment. On the other hand, it is possible to devise 
extremely lightweight recovery mechanisms with negligible 
runtime overheads under no-fault operating conditions
which is the focus of this paper.

In particular, we propose \shortnm, a lightweight and compiler-assisted 
technique which can repair a crashing application from its remaining 
uncorrupted state on-the-fly so that the application can continue the 
fault-free execution rather than being terminated and restarted with a 
checkpoint. Considering that the common use of ECC (Error-Correcting Code) 
can mask majority of transient errors in the memory of HPC machines, 
\shortnm mainly focuses on {\it those manifesting from CPU data paths} that 
are difficult or impractical to protect using ECC-like techniques and are 
attracting increasing concern in the HPC community. For example, \citet{OPDS2017} project 
that a hypothetical exa-scale machine built with $190,000$ cutting-edge Xeon 
Phi processors would experience daily transient errors with their memory areas 
protected with the ECC. 

\shortnm is motivated by an insight we observed from empirical instruction-level 
fault injection experiments, in which we adopt a model where the occurrence of 
a transient fault essentially causes an instruction to produce an incorrect 
result. While we present the details of this study in sections ~\ref{sec:inject-method} and~\ref{sec:trans-manifest}, we summarize 
its key findings here. The key result shows that the majority of soft failures 
manifest via hardware traps typically within a few dynamic instructions after a transient faults
an instruction. 
Specifically, as much as $99.08\%$ ($89.8\%$ on average) of these soft failures 
manifest themselves by causing a {\tt SIGSEGV}, because the fault corrupts 
the address calculations, thus leading to an invalid memory access. Majority of the HPC codes
involve very heavy array accesses in long running loops leading to complex address calculations. An example is shown
in Fig.~\ref{lst:loop}, which is a snippet extracted and simplified from 
GTC-P. For updating {\tt phism[i]}, a significant number of calculations 
are performed for computing array indexes for {\tt wtp}, {\tt phitmp} and 
{\tt jtp}. Hence, such array accesses stand a good chance of experiencing the impact 
of transient faults 
in practice. When a fault corrupts the computation of, e.g., {\tt k-1}, 
it would finally lead the application to access invalid address for 
{\tt wtp}\footnote{If the induced error is small enough, it may also lead 
application to access an incorrect element in {\tt wtp}, which could 
lead to SDCs, a topic which was covered in our other work~\cite{CGMS2018} and is outside
the scope of this paper which mainly focuses
on examining how a {\it crash} manifests from a transient fault, and how 
to recover from it.}, and crash the application by producing a 
{\tt SIGSEGV}, which can be essentially detected by OS for free. This 
zero-overhead detection of some manifestations of transient fault is crucial in that it allows the creation of a 
system that can potentially recover from some subset of transient faults, 
and improves the reliability of HPC systems without imposing a run-time 
overhead. In this simple case, we can recompute the index for {\tt wtp} 
by replaying the whole index computation {\tt (k-1) * 2 * n + i * 2}. 
Regardless which binary operation in this index computation is corrupted 
by a fault, redoing it will undoubtedly return the correct index, as 
long as initial values of {\tt k, n, i} are untainted and are available
in memory or registers. 
We call the corresponding instructions in the binary code an {\it RSI}
({\it Recoverable Sequence of Instruction}).

\begin{figure}[!tbp]
\begin{lstlisting}[language=C, style=c]
for (int k = 1; k < mzeta + 1; k++) {
  for (int i = start; i < end; i++) {
    j = k - 1;
    l = 2 * n;
    m = 2 * i;
    idx = j * l + m; // idx = (k-1)*2*n + i*m
    phism[i] = 0.25 * (1.0 - wtp[idx]) * phitmp[(mzeta + 1) * jtp[idx] + k];
    ......
  }
}
\end{lstlisting}
\caption{A snippet extracted from {\it GTC-P}.}
\label{lst:loop}
\end{figure}

Motivated by the above observation, \shortnm's approach for recovering 
from soft failures focuses on pre-building  a ``recovery kernel" 
$RK$ for each memory access instruction $I$ in the application. When $I$ 
is detected accessing an invalid address, $RK$ will be played to recompute 
the correct memory address for $I$. A ``recovery kernel" is similar to a 
function in C programming language. The body of the kernel is the RSI for 
the corresponding memory access instruction, and the parameters of the 
kernel are input values for the RSI. For illustration, Fig.~\ref{lst:kernel} 
shows the ``recovery kernel" for {\tt wtp} in Fig.~\ref{lst:loop}. When it
is executed for recovery, \shortnm will retrieve its parameter values from
the process address space at runtime. One of the key contribution 
of \shortnm is identifying the RSI and constructing the 
``recovery kernel" for each memory access instruction in applications. 
The concept of RSI plays a key role in this work; RSI dictates which values
are needed for replay and the availability of the values at recovery point dictates
if recovery is possible or not. Soon after the occurrence of a fault, the application
state continues to get modified. Empirically we observe that the fault leads to a crash
within a very small number of executing instructions. During this interval, fortunately a lot of
replay variables needed by RSI are not overwritten by the intervening instruction
execution and are still in-tact in terms of their original values. Such
values are replayed by RSI for recovery. In short, application crashes 
that occur because of transient faults in an RSI are always recoverable 
through the techniques presented here, and those outside them are not. 
The execution-weighted fraction of application instructions in RSIs 
determines the degree of our fault protection. However, identifying 
the RSI is not straight-forward. First, the code optimization and 
generation techniques may transform code in many ways, which would 
prevent us to construct the RSI by simply and aggressively cloning 
address computations. For the above example, the register assigned 
to {\tt n} may be reused by compiler to store the result of {\tt 2 * n}, 
making {\tt n} unavailable. In this case, it is impossible to 
correctly replay the recovery kernel for {\tt wtp} due to the lack 
of required value for {\tt n}. To this end, \shortnm employs the
live variable analysis to ensure that every parameter for the constructed 
kernel is accessible from the process address space at runtime. 
Second, loop index variables are essential components for accessing 
the array elements. If soft failures are due to the corruptions to their 
updates (e.g., {\tt i++}), their values should be recovered before 
replaying the address computation to successfully repair the soft failures.
For addressing this issue, \shortnm exploits side effects introduced 
by induction variable based code optimization techniques based on strength reduction and loop 
unrolling, which are widely adopted by modern compilers. While these 
code optimization techniques were mainly designed to improve the 
execution speed, they introduce equivalent computation patterns 
and values (semi-redundancies) into the code. Those semi-redundancies 
are in the form of sets of state elements which are updated synchronously 
across loop iterations, a situation allowing a corruption in one of those 
elements to be potentially repaired by inferring its proper value from the 
uncorrupted value of another in that set. \shortnm exploits this observation 
to augment recovery kernels such that they can repair corruptions in one 
induction variable by referencing the uncorrupted value in another induction 
variable (that synchronously updates with it from one iteration to the next) from the same code region.

\begin{figure}[!tbp]
\begin{lstlisting}[language=C, style=c]
// the recovery kernel for wtp1 in Fig. 1
uint64_t recovery_kernel(int *wtp1, int k, int n, int i) {
    return (wtp + ((k - 1) * 2 * n + i * 2));
}
\end{lstlisting}
\vspace{-.1in}
\caption{A sample recovery kernel.}
\label{lst:kernel}
\vspace{-.1in}
\end{figure}

\shortnm is designed with two components: a front-end, which is consists
of a set of compiler passes for detecting a set of synchronously updating sets of
induction variables and constructing
aforementioned ``recovery kernels", as well as a runtime system for performing
actual recovery services. To minimize the overheads, recovery kernels for an 
application are compiled into a stand-alone shared library, which is loaded 
dynamically by the runtime when a crash-causing error is experienced. The runtime
is essentially a signal handler for {\tt SIGSEGV}. It is invoked only upon a 
failure to diagnose which instruction caused the invalid memory access, and
disassemble the instruction to determine which operand is referring to 
a memory address. Based on the address of the instruction, the runtime will 
then search, load and execute the related recovery kernel to recompute the 
accessed memory address for the instruction, and update the related 
operand. The runtime is designed to be transparent to 
applications and requires no instrumentation or modification to 
applications' source code. It is implemented as a shared library 
that can be automatically loaded through setting the {\it LD\_PRELOAD} 
environment variable. Because the runtime is not activated unless a 
crash-causing fault occurs, the small load-time overhead of installing 
a signal handler and the tiny memory overhead for storing the signal 
handler are its only impact on an application's execution under application execution
with no-fault.

In summary, this paper makes the following major contributions:
\begin{itemize}
    \item We studied the manifestation of soft failures 
    in modern scientific applications through empirical instruction-level 
    fault injection experiments ~\ref{sec:inject-method} and ~\ref{sec:trans-manifest}. 
    We classified these soft failures based on hardware 
    trap symptoms, and examined their manifestation latency measured in terms 
    of number of dynamic instructions. The study pointed to a direction that one could
    devise recovery kernels that recompute the array offsets by leveraging available state. 
    
    \item We propose \shortnm, a new failure recovery strategy 
    for scientific applications to survive soft failures. \shortnm 
    leverages hardware detection of memory access violations to 
    repair crashed architecture states on-the-fly by replaying 
    computations that are extracted and cloned from applications. 
    \shortnm also exploit the properties of modern code optimization 
    techniques for resilience purpose. \shortnm is lightweight. 
    Except requiring some offline code analysis effort for building 
    recovery kernels, \shortnm incurs negligible (if not {\bf zero}) 
    runtime overheads and tiny memory overheads during the 
    normal run of applications.
    
    \item We design and implement \shortnm based on the LLVM 
    framework and the Linux system. While more engineering work is 
    needed to support -O2/-O3 optimizations, our prototype of 
    \shortnm is a solid step towards a lightweight resilience 
    mechanism for soft failures.
    
    \item We evaluat \shortnm with 4 scientific workloads
    and the NPB benchmark suite. The 
    results show that, on average, \shortnm can recover about 
    $84\%$ of soft failures for the evaluated workloads within 
    dozens of milliseconds, allowing parallel applications to 
    finish their jobs with almost no delays even when crash-causing 
    errors happen during their execution. 
    % We also present 
    % preliminary evaluation results for \textit{BLAS}, showing 
    % that \shortnm can support for failure recoveries in libraries 
    % with a high coverage rate and negligible performance hit.
    
\end{itemize}

The rest of paper is organized as follows: 
Section~\ref{sec:motivation} introduces the background for \shortnm; 
Section~\ref{sec:design} and Section~\ref{sec:impl} 
present the overall framework including detailed algorithms and implementation details of \shortnm respectively. 
Next, evaluation results are presented in
Section~\ref{sec:eval}, and the related state-of-the-art 
is discussed in 
Section~\ref{sec:relate}. Finally, we present our 
conclusion in Section~\ref{sec:conclusion}.
\section{Background}\label{sec:motivation}
In this section, we will briefly introduce the compiler techniques 
that are used by the techniques introduced in the paper. 
We will first introduce live variable analysis, which is critical for 
building recovery kernels, and then present induction variable based code optimization 
techniques, including strength reduction and loop unrolling, with 
a focus on how they produce semi-redundancies that can be exploited 
for resilience purpose with a simple example.

\subsection{Live Variable Analysis}
Live variable analysis (or simply liveness analysis) is a classic 
data-flow analysis in compiler for calculating the variables that 
are live at each point in the program. A variable is live at some 
program point {\it p} if its value is used along some control flow path
that emanates from {\it p}, which means the 
variable may be read before the next time it is written. Otherwise, 
the variable is dead at the program point. A live variable is a candidate
for being allocated in a register. Consider the snippet in 
Fig.~\ref{lst:loop}, the set of live variables between lines 6 and 
7 are \{{\tt wtp}, {\tt idx}, {\tt phitmp}, {\tt mzeta}, {\tt k}\}, 
because all of them are used in line 7; and {\tt j}, {\tt l}, 
{\tt m} are dead if they are not used after line 7. If a variable 
is live at a specific program point, the compiler will preserve its 
value somewhere, e.g, in a register or spill to a stack, during the
process of register allocation and code generation. If a variable 
is dead, the compiler can reuse the register assigned to it without 
the need of saving its value. \shortnm leverages this analysis to 
guarantee that, for every recovery kernel, it always has access to 
its parameters at runtime by ensuring that every parameter of the 
kernel is live at the corresponding memory access instruction, i.e., 
its current definition can be found in either a register or a spill memory
location allocated on the stack. In our analysis, we use LLVM's SSA-based representation in which each definition of
a variable is identified through a unique name. 

\begin{figure}
    \centering
  \begin{subfigure}{\linewidth}
    \centering
\begin{lstlisting}[language=C, style=c]
c = 7;
for (i = 0; i < N; i++) {
    y[i] = c * i;
}
\end{lstlisting}
    \vspace{-.1in}
    \caption{Original Example Code}
    \label{lst:orig}
  \end{subfigure}
    
  \vspace{.15in}
  
  \begin{subfigure}{\linewidth}
    \centering
\begin{lstlisting}[language=C, style=c]
c = 7, k = 0;
for (i = 0; i < N; i++) {
    y[i] = k;
    k = k + c;
}
\end{lstlisting}
    \vspace{-.1in}
    \caption{Transformed code using Strength Reduction}
    \label{lst:sr}
  \end{subfigure}
  
  \vspace{.15in}
  
  \begin{subfigure}{\linewidth}
    \centering
\begin{lstlisting}[language=C, style=c]
c = 7;
for (i = 0; i < N; i+=2) { // assume N%2 = 0
    y[i] = c * i;
    y[i+1] = c * (i + 1);
}
\end{lstlisting}
    \vspace{-.1in}
    % \caption{Transformed code using Loop Unrolling (For simplicity of 
    %     illustration, here we assume N \% = 2. 
    %     More codes were generated to handle corner 
    %     case in production compilers)}
    \caption{Transformed code using Loop Unrolling. }
    \label{lst:lu}
  \end{subfigure}  
  \caption{Semi-redundancy introduced by code optimizations}
  \vspace{-.15in}
\end{figure}

\subsection{Strength Reduction}
Strength reduction is a code transformation technique 
in modern compilers that replaces certain costly instructions 
with less expensive ones without changing programs' correctness. 
The classic example of strength reduction is to convert expensive 
multiplications into left shifts. Although strength reduction 
is a global optimization, it is typically applied to computations in 
loops, since most of a program's execution time is typically spent in 
a small section of code which is often inside loops that is executed 
over and over. Incidentally, this portion of code is also more highly likely to 
experience transient errors. Strength reduction looks for expressions 
involving an induction variable (a value which is 
changing by a known amount in each iteration of the loop) and transform calculations
based on it into lesser expensive counterparts. If applicable, 
strength reduction will transform these expressions into an equivalent but 
more efficient form. For illustration consider Fig.~\ref{lst:sr} which shows the 
transformed code after applying strength reduction on the code in 
Fig.~\ref{lst:orig}. As shown in the figure, the original multiplication 
operation \texttt{c * i} is replaced with (reduced to) a cheaper 
addition operation \texttt{k + c}, so the performance of the code 
is improved. 
However what's important for \shortnm is that the introduced new 
expression {\tt k + c} shares a similar computation pattern to 
{\tt i++}. This provides an opportunity to recover the value of 
{\tt i}, if it is corrupted, by referring to {\tt k} as long as 
the initial and step values of these two variables and their 
updates are available. 
In particular, the correct value for {\tt i} can be recomputed 
as {\tt i = k / c} if {\tt k} is in-tainted
(The initial values for {\tt i} and {\tt k} are $0$, and their 
step sizes are $1$ and $c$ respectively).

\subsection{Loop Unrolling}
In addition to strength reduction, loop unrolling is another compiler 
optimization technique that could introduce semi-redundancies to codes. 
The main goal of loop unrolling is to increase a program's speed by 
reducing (or eliminating) instructions that control the loop (such 
as end-of-loop tests on each iteration), reducing branch penalties, 
and hiding latency (e.g., the delay in reading data from memory) through better pipelining etc. 
To eliminate these computational overheads, loop unrolling re-writes 
the loop as a repeated sequence of similar independent statements. 
Fig.~\ref{lst:lu} shows the transformed code after applying loop 
unrolling on the code in Fig~\ref{lst:orig} by unfolding the loop 
body twice. the transformation reduces the number of end-of-loop 
tests by almost half in the new code. Meanwhile, it also introduces
two computing operations that are based on induction variable
copies of i, such as \texttt{i + 2} and 
\texttt{i + 1}. If one of the copies and the address calculation based on it
is corrupted due to a fault, a second copy that is value related is
available in the same loop body for recovery. Some register allocation techniques
such as coalescing might try to mangle the two copies to reduce register pressure;
by selectively disabling coalescing, the resilience improves with a negligible
performance impact.

\section{Recovery Framework}\label{sec:design}

Based on the empirical observation of very short interval between incidence of a transient error and its manifestation in which the state of recovery parameters is in-tact, we designed the compiler based \shortnm 
environment to focus on recovery from {\it SIGSEGV} faults. 
In this section, we first depict the overall architecture of 
\shortnm, and then dive into the design details of each component.

\subsection{Overview}
\begin{figure}[!tbp]
    \centering
    \includegraphics[width=\columnwidth]{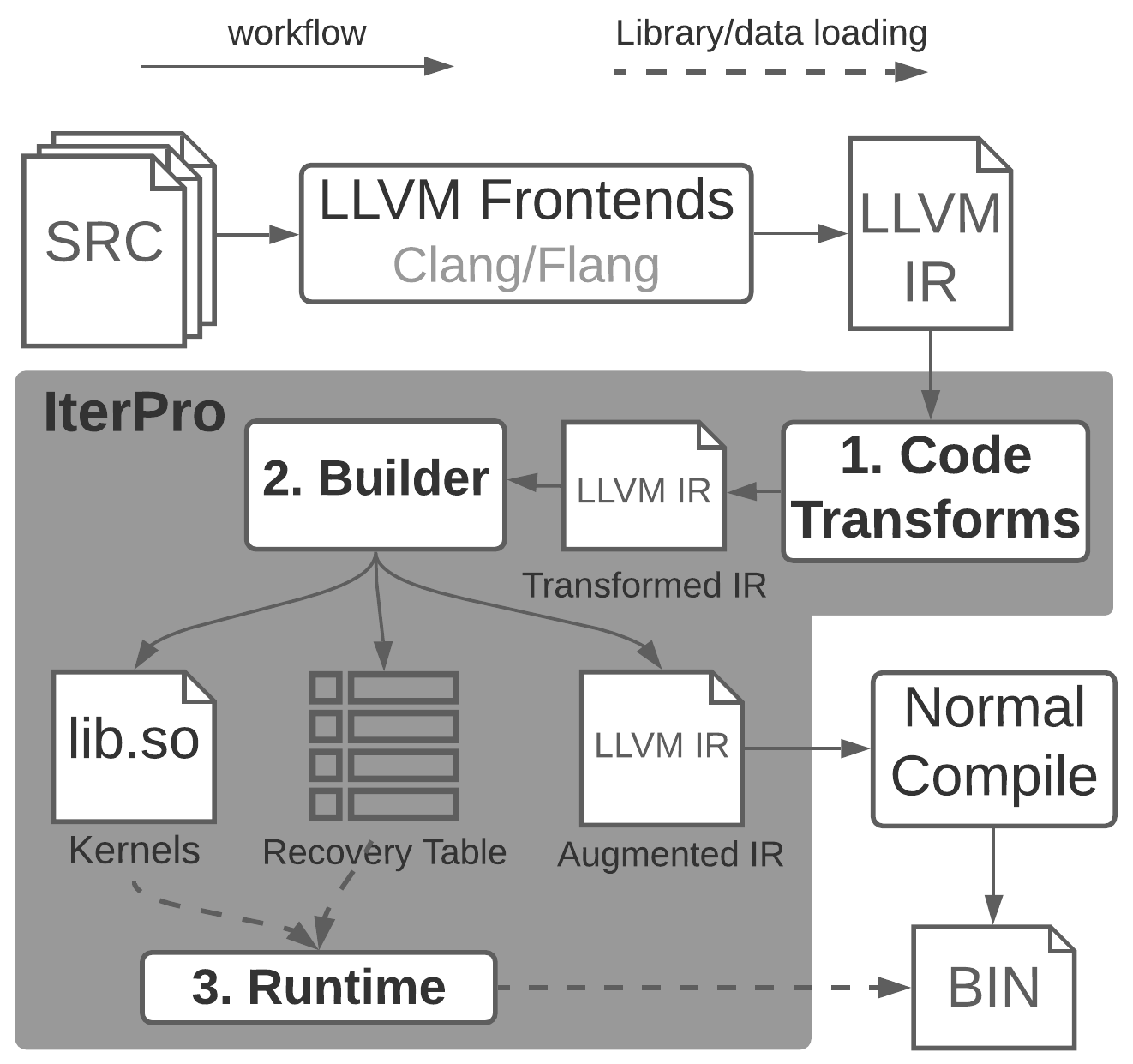}
    \caption{Overall architecture of \shortnm}
    \label{fig:arch}
\end{figure}

\shortnm is a compiler-assisted failure recovery mechanism to recover 
impacted scientific applications from transient-error-induced crashes
with (almost) zero runtime overheads, such that applications can continue
their executions as normal, instead of being terminated and restarted 
with expensive checkpoint-restart mechanism. 

Fig.~\ref{fig:arch} shows the workflow of \shortnm. It works on LLVM IR, 
a light-weight low-level intermediate representation of programs. There are 
many LLVM front-ends, such as Clang, Flang, and 
DragonEgg, that compile applications written in different 
programming languages into LLVM IR. Therefore, working on LLVM IR 
allows us to \shortnm support a majority of scientific applications written in 
C, C++ or FORTRAN. To explicitly expose the side effects introduced by 
modern code optimization techniques for recovery of induction variables, 
the original LLVM IR code is first transformed with a new \textbf{code transform} pass, 
and then the \armor (another compiler pass) is invoked to build recovery 
kernels for all memory access instructions (one recovery kernel per memory
access instruction) and induction variables if applicable. To minimize the 
overhead of \shortnm, recovery kernels are compiled into a stand-alone shared 
library, which is loaded only when an error is encountered. In addition to 
recovery kernels, the \armor also generates a {\bf Recovery Table} and augments 
the application's LLVM IR with appropriate debug data. The {\bf Recovery Table} 
and the debug data together provide information to the \safeguard about how 
to access and execute a recovery kernel. Upon a invalid memory access error, 
\safeguard will be activated to diagnose the error, find the appropriate 
recovery kernel and retrieve related parameter values from the process 
address space with the help of the recovery table and debug data, and finally 
execute it to repair the corrupted architecture state. The \safeguard itself 
is designed and implemented as a shared library as well. It will be automatically 
loaded by setting the {\it LD\_PRELOAD} environment variable, requiring no code 
changes to applications. Basically, it overloads the default {\it SIGSEGV} 
signal handler of applications with a customized one. Besides such initialization 
work, the \safeguard is not activated unless an invalid memory 
access is detected. Such light-weight design makes \shortnm incur almost 
negligible overheads during the normal execution of applications. 

Although \shortnm is a complex system due to the nature of the problem 
it aims to address, in the rest of the section, we will mainly focus on 
the novel idea of leveraging the side effects introduced by modern code 
optimization techniques for resilience purposes, which is completely new 
as compared to other studies. The design details for \armor, {\bf Recovery Table} 
and \safeguard have been presented in our conference paper~\cite{CGSQ2019}. 
Hence, only a brief introduction to these components will be presented for 
the sake of completeness of the paper. 
% Please refer to the conference paper 
% for details.

\subsection{Recovery for Induction Variables}\label{subsec:iv}
The philosophy behind \shortnm for recovery of induction variables 
is pretty straightforward. For a given induction variable \texttt{i}, 
updated as \texttt{i = i + s$_i$}, \shortnm will leverage scalar-evolution 
analysis to find another induction variable(s) {\tt k} in the same code region, 
which is a loop, such that {\tt k} is updated with a computation pattern 
(\texttt{k = k + s$_k$}) similar to {\tt i} and {\tt k} is not used with 
{\tt i} at the same time to compute a memory address (e.g., {\tt y[i+k]}). 
And {\tt k} is then considered as a partner (or co-related induction variable) 
to {\tt i}, such that if {\tt i} is corrupted by a fault, \shortnm is
able to recover it by referring to {\tt k} (vice versa) based on the following 
equation:
\begin{equation}
    i = \frac{k-k_0}{s_k} \times s_i + i_0
    \label{eq:pair}
\end{equation}
where, $i_0$ and $k_0$ are initial values of $i$ and $k$ respectively.

While it would very difficult to find such computation pairs in original 
source codes, the code optimization techniques deployed in modern compilers, 
such as strength reduction and loop unrolling, introduce more 
opportunities in transformed codes (See section~\ref{sec:motivation}), 
which are only accessible by compiler passes at the IR level. To be able to successfully 
recover {\tt i} when it is corrupted, \shortnm must know or have accesses 
to initial values of {\tt i} and {\tt k} and their step sizes at runtime.
In other words, when {\tt i} is corrupted, \shortnm should be able to: 
1). find its partner {\tt k}; 2) their initial values {\tt i$_0$} and 
{\tt k$_0$}; 3) their step sizes {\tt s$_i$} and {\tt s$_k$}; and 4)
the current value of {\tt k}. If these values are not compile-time constants,  
\shortnm must ensure that they are stored either in a register 
or on stack during the code generation pass, and such that they are available 
(the location storing them is not reused by others) regardless when
they are accessed during runtime. Unfortunately the semi-redundancies
introduced by the aforementioned code optimization techniques might be 
not directly exploitable for resilience purposes due to following challenges: 
\begin{enumerate}
    \item No partner is exposed in the IR. In such case, even though these techniques 
    introduced semi-redundancies, but they don't introduce new variables. 
    An example is shown in Fig.~\ref{lst:lu} where {\tt i + 1} shares a
    similar computation pattern to {\tt i += 2}. However, it is useless 
    since they both depend on {\tt i}. In particular, if accessing to 
    {\tt y[i]} failed because of a fault in {\tt i}, there is no partner 
    available for \shortnm to recover it. It may be possible that two new
    temporaries are generated at symbolic level; however, they could be coalesced
    into one register during the allocation phase. 
    
    \item Sometimes initial values or step sizes are not available at runtime when
    a failure is detected. This typically happens to pointer variables as illustrated 
    in Fig.~\ref{lst:intuitives}. In this case, \shortnm would be able to find the 
    partner for {\tt i}, which is {\tt A}, but it may fail to find its initial value 
    {\tt A$_0$}. This is because the code generator typically maps {\tt A} to a register, 
    saying {\tt \%rax}, and updates it in-place simply with {\tt add \%rax, 8}. 
    Therefore, the initial value for {\tt A} is not preserved in applications' 
    process address spaces. 
\end{enumerate}

\begin{figure}[!btp]
\begin{lstlisting}[language=C, style=c]
for (i = 0; i < N; i++) {
    sum += *(A++); // from sum += A[i];
    sum /= (B[i+1] + C[i-1]);
}
\end{lstlisting}
\vspace{-.1in}
\caption{A sample example.}
\label{lst:intuitives}
\vspace{-.15in}
\end{figure}

In order to address these problems, \shortnm introduces two additional code 
transformations, named independent compute promotion (ICP) and micro-checkpoint generation. 
For the first case, \shortnm leverages ICP to transform dependant 
computations into independent ones, if possible, by introducing new 
variables and/or by disabling register coalescing. And for the second case related to the loss of initial values, \shortnm 
introduces code to store (checkpoint) related initial values in the stack, 
such that they are always available when they are needed for recovering 
corrupted induction variables.

\begin{figure}
\begin{subfigure}{\linewidth}
    \centering
\begin{lstlisting}[language=C, style=c]
c = 7;
for (i = 0, k=1; i < N; i+=2, k+=2) {
    y[i] = c * i;
    y[k] = c * k;
}
\end{lstlisting}
    \vspace{-.05in}
    \caption{Independent code promotion}
    \label{lst:ip}
\end{subfigure} 
\vspace{.15in}

\begin{subfigure}{\linewidth}
    \centering
\begin{lstlisting}[language=C, style=c]
S = A;
B = S;
for (i = 0; i < N; i++) {
    sum += *(B++); // from sum += A[i];
}
\end{lstlisting}
    \vspace{-.05in}
    \caption{Micro-checkpoint}
    \label{lst:enforce}
\end{subfigure} 
% \vspace{-.05in}

\caption{Code Transformations in \shortnm. C/C++ are used for illustration only. 
\shortnm actually works on LLVM IR code.}
% \vspace{-.05in}
\end{figure}

\subsubsection{Independent Compute Promotion} 
Typically, semi-redundancies introduced by loop-unrolling exhibiting in the 
code in form of \emph{derived induction values} (e.g., {\tt i + 1} in Fig.~\ref{lst:lu}). 
Per discussion before, such semi-redundancies can't be directly exploited by \shortnm, 
so we introduce compiler pass, ICP, which transforms these derived induction values 
into independent computations. It will create new induction variables along with their related update
instructions to replace original derived induction values. For illustration, 
Fig.~\ref{lst:ip} shows the transformed code derived the code in Fig.~\ref{lst:lu},
in which a new variable {\tt k} is created and original {\tt i + 1} is replaced 
with {\tt k} and {\tt k + 2}. In particular, note that {\tt k} is completely independent from {\tt i}, 
therefore they can be inferred to recover each other if either one is corrupted. 
It is worthwhile to note that while ICP does demand an additional register, 
it doesn't introduce new computation. Such change is often hidden in superscalar 
processors. Hence, it has negligible penalties to applications' performance.  

Algorithm~\ref{alg:icp} shows the core steps of independent compute promotion.
For each loop in LLVM IR codes, ICP iterates over each binary operator in the 
loop. For those who are directly used (both directly and indirectly) in address 
computations, \shortnm will create new induction variables to replace them, if 
they can be expressed in form of ({\tt i = i + s}) based on scalar-evolution analysis, 
where {\tt s} is a loop invariant value (it doesn't need to be a constant). 

\begin{algorithm}[!tbp]
\caption{The Pseudo Code for Independent Compute Promotion.}
\label{alg:icp}
\begin{algorithmic}
\Function{doIndependencePromotion}{{\tt loop}}
\For{every binary operator $BO$ in {\tt loop}}
    \State $Expr \gets$ \Call{getSCEVExpr}{$BO$}
    \State $isAddRec \gets \Call{isSCEVAddRecExpr}{Expr}$
    \State $isInAddr \gets$ \Call{isUseInAddrCompute}{$BO$}
    \If {${\it isAddRec}$ \&\& ${\it isUsedInAddr}$}
        \State $initVal \gets \Call{getStartValue}{Expr}$
        \State $stepVal \gets \Call{getStepValue}{Expr}$
        \State $IndPhi \gets \Call{createPHINode}{initVal}$
        \State $Inc \gets \Call{createIncOp}{IndPHI, stepVal}$
        \State $IndPhi \rightarrow \Call{addIncomingValue}{Inc}$
        \State $ BO \rightarrow \Call{replaceUsesWith}{IndPhi}$

    \EndIf
\EndFor
\EndFunction
\end{algorithmic}
\end{algorithm}

\begin{algorithm}[!tbp]
\caption{The pseudo code for micro-checkpoint}
\label{alg:chkpt}
\begin{algorithmic}[]

\Function{doCheckpoints}{{\tt loop}}
\For{every induction variable $IV$ in {{\tt loop}}}
    \State $Latch \gets \Call{getLoopLatch}{\tt loop}$
    \State $Init \gets \Call{getStartValue}{IV}$
    \State $Const \gets \Call{isConstant}{Init}$
    \State $Live \gets \Call{isLiveAt}{Init, Latch}$
    \If{!${\it Const}$ \&\& !${\it Live}$}
    \State $Var \gets \Call{createLocalVariable}{}$
    \State \Call{createStore}{Var, {\tt IV}}
    \State $Val \gets \Call{createLoad}{Var}$
    \State {\tt IV} $\rightarrow \Call{repalceAllUsesWith}{Val}$
    \EndIf

\EndFor
\EndFunction
\end{algorithmic}
\end{algorithm}
\vspace{-.1in}

\subsubsection{Micro-checkpoint}
Micro-checkpoint is applied only to induction variables whose initial values 
are not live across the loop body. If a value is not live across the loop body, 
the location for holding this value could be reused by other variables at runtime,
which means it could be not accessible by the recovery mechanism. 
For these induction variables, \shortnm will checkpoint 
their initial values into the stack frame by creating new local variables and inserting a store instruction.
The transformed code for the code in Fig.~\ref{lst:intuitives} is shown 
in Fig.~\ref{lst:enforce}, in which a new local variable {\tt S} is 
allocated to store the initial value (base address) of {\tt A}. And a 
new variable {\tt B} is introduced as an alias to {\tt A} to iterate 
over elements in the array. And {\tt B} will be identified as the 
partner to {\tt i}. While {\tt B = S} looks redundant, but it is 
not trivial. It provides \shortnm heuristics about where to find 
initial values for {\tt B}. Notably, the new code has substantially similar 
performance as the original code, since the instruction insertions are {\bf outside} the loop body. 
The pseudo code for micro-checkpoint is shown in Algorithm~\ref{alg:chkpt}.
It iterates over each induction variable of a loop. If {\tt init} is not a 
constant number, and it is not live (based on liveness analysis) at the end 
of corresponding loop, \shortnm will then create a new local variable on 
the stack to store its initial value. 

Please note that although C/C++ syntax is used with above examples for the sake of clarity, \shortnm and the above algorithms
actually operate 
on LLVM IR.

\subsection{Building Recovery Kernels} 
\armor is a compiler pass working on LLVM IR, in which memory accesses are 
issued explicitly through either {\it LoadInst} or {\it StoreInst} instructions. 
For each memory-access instruction $I$, except those directly accessing an 
static memory location, e.g., a local variable in the stack or a global 
variable~\footnote{these memory accesses don't have any address computations 
associated with them.},
\armor starts from its address operand {\tt ad-op} and works backwards to 
determine its RSI by iteratively including every value and its corresponding 
instruction in the def-use chain of {\tt ad-op}, until it meets a predefined 
set of {\bf Terminal Values}. That is the backward transitive closure for 
determining the RSI which is not extended beyond Terminal Values. 
For a memory access instruction $I$, a \textbf{terminal value} is a LLVM IR 
instruction/value which is live at $I$, with at least one of its 
operands being dead at $I$ and the dead operand being not computable
from other live instructions/values.
\armor treats the \textbf{terminal values} as the parameters of the recovery kernel, 
and clones all other checked instructions into the function body of the kernel. 
 In addition to the above definition of \textbf{terminal values}, \textit{AllocaInst}s, 
\textit{GlobalVariable}s, \textit{Argument}s, and {\it PHINode}s 
(typically representing induction variables) are also treated as 
{\bf terminal values} too. Please refer to~\cite{CGSQ2019} for the 
details of this building process including an illustrative example (section 3.2). The intuition behind {\bf terminal values} 
is that they are guaranteed to be found in the process address space 
at runtime, when the corresponding kernel is executed to repair a crash. 
In addition, for each induction variable, additional recovery codes are 
generated by \shortnm based on the technique introduced in~\ref{subsec:iv}, 
which is not covered in~\cite{CGSQ2019}. 
Meanwhile, \armor also attaches a unique debug 
data in tuple of ($file$, $line$, $column$) for each memory access 
instruction. The debug data will be finally embedded in the final 
binary code of the application, and serve as the key to find the 
recovery kernel for the memory access instruction. It may be noted that \shortnm doesn't 
require the real debug data of the program, since it won't map 
instructions to original source-code statements.

\subsection{Recovery Table}
{\bf Recovery Table} is an important metadata generated by 
\armor to describe recovery kernels for the \safeguard. It 
contains information about how to access a recovery kernel 
and which are the parameters to the kernel, and plays an 
important role in providing synchronization between \armor and \safeguard, 
which work on different representations of applications (\armor
works on IR representation, and \safeguard works on binary code).
{\bf Recovery Table} is simply a key-value table as shown in 
Table~\ref{tab:rktble}. For each recovery kernel, \armor will 
register an entry for it in the table with three pieces of information: 
\begin{itemize}
    \item {\bf key}, which uniquely represents the corresponding 
    memory access instruction. \armor uses the aforementioned debug 
    data for this purpose.
    
    \item {\bf symbol}, which represents a recovery kernel. It is 
    simply the function name of the recovery kernel. 
    
    \item {\bf parameters}, which describes the parameters of the kernel. 
    They are simply the variable names. For each parameter, \armor will 
    create a variable description debug entry, for which the debug information 
    subsystem of the compiler will automatically generate a debug information 
    entry (DIE) to describe the variable in machine code, which will be used 
    for determining where to retrieve the parameter values. 
\end{itemize}

\begin{table}[!tbp]
    \centering
    \caption{Recovery table for describing recovery kernels}
    \label{tab:rktble}
    \vspace{-.05in}
    \begin{tabular}{|c|p{5cm}|c|}
        \hline
        key  & symbol            & parameters       \\ \hline
        key1 & recovery\_k1(int16, int, int) & $a$, $b$, $c$   \\ \hline
        key2 & recovery\_k2(float, int32) & $m$, $n$        \\ \hline
        key3 & recovery\_k3(int8, int64)  & $d$, $e$        \\ \hline
    \end{tabular}
\end{table}

\subsection{Runtime System}
The \safeguard system of \shortnm is basically a customized signal handler
for {\it SIGSEGV} faults. It overrides the default signal handler for 
{\it SIGSEGV} immediately after the process is started by leveraging the 
``constructor" attribute in modern compilers, and will be automatically 
activated by the operating system upon a {\it SIGSEGV} fault. To repair
the fault, it first finds the corresponding recovery kernel based on the 
address of the instruction issuing the {\it SIGSEGV} signal, retrieves 
its parameter values from the process address space, and then executes 
the kernel to recompute the accessed memory address. If the kernel-computed 
address is the same with the one accessed by the instruction (which is a malformed
address due to the transient), the \safeguard
system will abort the recovery, leading the application to be terminated by 
the OS. Otherwise, it will fix the corrupted architecture state based on the
replay of the RSI leading to repaired array access. Since the \safeguard is not 
activated until a {\it SIGSEGV} fault occurs, it has almost {\bf zero} runtime 
overhead during the normal execution of applications. To minimize the memory 
overhead, the \safeguard only loads recovery kernels when it is activated, 
and releases the related memory immediately after finishing its job. 
\section{Prototype}\label{sec:impl}
We implemented a prototype of \shortnm on X86\_64 platform and Linux 
OS. The compiler passes, including {\bf code transforms} and \armor 
are implemented based on LLVM 6.0.1. \armor treated some LLVM 
\textit{CallInst}s as a normal binary operators, if they simply 
call mathematical kernels, e.g, \textit{sqrt}, or user-implemented 
functions that don't update global variables and arguments. It doesn't 
clone the implementation of these callee functions, hence, when the 
recovery kernels are compiled into a shared library, it is necessary 
to build them with binary source files containing the user-implemented 
simple functions, and link them with necessary libraries. For the \safeguard 
system, it leverages libdwarf library to read the 
debug data and the libffi library to execute calls 
to the recovery kernel. Since ``ffi\_call" takes pointers as arguments, 
the address of a variable, instead of a value, is retrieved from the 
process space. Finally, {\bf recovery table} is implemented 
based on google protobuf-3.6.0, and the MD5 hash 
of the debug information tuple $(file, line, column)$ is computed
with the mhash library and used as the key.
\section{Evaluation}\label{sec:eval}
% This section presents evaluation results of \shortnm. 
In section, we will first introduce the evaluation 
methodology, the fault model, the results of our fault-injection experiments and then present evaluation results for \shortnm.

\begin{table*}[!ht]
    \centering
    \caption{Scientific workloads from different scientific domains and implementing different algorithms}
    \label{tab:sci-apps}
    \vspace{-.05in}
    \begin{tabularx}{\textwidth}{|c|c|X|}
      \hline
              Workload        &       Language       & Description        \\ \hline
              
     \multirow{2}{*}{GTC-P}  & \multirow{2}{*}{C}  & A 2D domain decomposition version of the GTC global gyrokinetic 
                                                     PIC code for studying micro-turbulent core transport. It solves 
                                                     the global, nonlinear gyrokinetic equation using the particle-in-cell 
                                                     method.                                                        \\ \hline
                                                     
      \multirow{1}{*}{HPCCG}  & \multirow{1}{*}{C++} & A simple conjugate gradient benchmark code for a 3D chimney 
                                                       domain on an arbitrary number of processors.                 \\ \hline
                                                     
    %   \multirow{2}{*}{miniMD} & \multirow{2}{*}{C++} & A simple, parallel molecular dynamics (MD) code. It 
    %                                                   performs parallel molecular dynamics simulation of a 
    %                                                   Lennard-Jones or a EAM system                                \\ \hline
      \multirow{2}{*}{CoMD} & \multirow{2}{*}{C} & A reference implementation of typical classical molecular 
                                                    dynamics algorithms and workloads as used in materials science. 
                                                         \\ \hline
        
      \multirow{2}{*}{miniMD} & \multirow{2}{*}{C++} & A simple, parallel molecular dynamics (MD) code. It 
                                                       performs parallel molecular dynamics simulation of a 
                                                       Lennard-Jones or a EAM system                                \\ \hline
        
      \multirow{2}{*}{NPB} & \multirow{2}{*}{C} & The NAS Parallel Benchmark (NPB) suite is a small set of programs 
                                                    derived from computational fluid dynamics (CFD) applications. It 
                                                    consists of $5$ kernels and $3$ pseudo-applications. In this work,
                                                    NPB3.0-C version is used.
                                                    \\ \hline
    \end{tabularx}
    \vspace{-.1in}
\end{table*}

\subsection{Injection Methodology}\label{sec:inject-method}
We evaluated \shortnm on an X86\_64 platform equipped with $48$ cores and $128$GB 
of memory, using empirical fault injection experiments, which were widely used in 
prior studies~\cite{LiVY2012,CGBM2014,AGKD2015,CSOG2017}. Similar to these studies, 
\shortnm focuses on faults in the CPU logic, assuming memory regions are protected 
with other techniques, such as ECC. The injection tool introduced in~\cite{CGSQ2019} 
is used in this work. To emulate the the impact of transient faults from the CPU logic, 
it injects a fault to the ``destination" operand of a randomly selected instruction 
right after the instruction is executed. Then the execution of the process is continued. 
A ``destination" operand is one of architecture states, e.g. a register, or a memory 
cell, that is updated by the instruction. For instructions having implicit destination 
operand(s), such as X86 {\tt idiv} $\%ecx$ which divides the value in $\%edx:\%ead$ by 
$\%ecx$ and store results in $\%eax$ and reminder in $\%edx$, one of the implied 
destinations, e,g, $\%eax$, is selected. To achieve this purpose, we first profiled 
the number of executions for each static instruction (from applications only) using 
the Intel Pin tool. Then we randomly select a static instruction for injection based 
on the numerical distribution of their executions, and also generate a random number 
based on the executions of the selected instruction to determine the program point at 
which the fault would be injected at runtime. In other words, a dynamic instruction is 
approximately represented by a pair $(I, n)$, which means the fault will be injected to 
the instruction $I$ after it is executed $n$ times. For each run of an application, only 
one injection is performed. The single-bit-flip fault model, which is widely used in 
previous studies~\citep{CSOG2017,FGDP2017}, is used in this work, which means, for 
each injection, it randomly flips a bit in the destination operand. We are particularly 
interested in faults that lead to process crashes and specifically how many of them 
are successfully recovered by \shortnm. For each workload, we performed \num{5000} 
injections. In all, we get $788\sim2791$ (depends on applications) injections that lead 
to process crashes. Then these injections were replayed to actually evaluate the 
performance of \shortnm.

Four scientific proxy applications, including GTC-P, HPCCG, CoMD and miniMD, as 
well as $8$ benchmarks from the NPB benchmark suite were used in our experiments. 
Table~\ref{tab:sci-apps} briefly presents their properties. These benchmarks are 
derived from production scientific applications for evaluating system performance. 
They contain compute-intensive kernels which typically dominate the execution of 
production scientific applications. Therefore, in production applications, these 
portions of codes are more likely to experience transient faults. All of these 
codes were compiled into LLVM IR codes with clang using the ``-O1'' flag.

\subsection{Manifestation of Soft Failures}\label{sec:trans-manifest}
In this subsection, we present our study about how transient errors 
manifest into crashes, which we think is the key to building efficient 
resiliency mechanisms. In particular, the results of this study inspired 
the design of \shortnm. While several recent papers~\cite{LiVY2012,CGBM2014,CSOG2017} 
have experimentally studied the impact of transient errors on scientific applications, 
none of them provided quite the insights necessary for devising efficient recovery 
mechanisms. In their studies, applications are treated as black-boxes, thus they
failed to provide adequate information about how transient errors manifest and 
propagate inside applications, which is critical for building efficient resiliency 
mechanisms. To fill this gap, we performed empirical fault injection experiments 
on four proxy scientific workloads in Table~\ref{tab:sci-apps}, and studied how 
(some of) the injected faults manifest, propagate and finally lead the application 
to crash by tracking the propagation of faults from instruction to instruction. In 
this study, we are specially interested in: 1) determining the major causes/symptoms 
of crashes; and 2) the latency of their manifestation in terms of number of 
instructions executed from the injection point to the crash point. To get solid and 
unbiased results, we performed \num{10000} injections for each workload.  

\begin{table}[!tbp]
    \centering
    \caption{The overall outcomes of fault injections}
    \label{tab:outcomes}
    \vspace{-.05in}
    \resizebox{\columnwidth}{!}{
    \begin{tabular}{|c|c|c|c|c|c|}
        \hline
        Workloads  &  Benign    & Crash         &     SDC     & Hang     \\ \hline
        HPCCG      & \num{3118} & \num{3409}    & \num{3472}  & \num{0}  \\ \hline
        CoMD       & \num{6433} & \num{2439}    & \num{1120}  & \num{8}  \\ \hline
        miniMD     & \num{951}  & \num{4065}    & \num{4984}  & \num{0}  \\ \hline
        GTC-P      & \num{6875} & \num{1644}    & \num{1479}  & \num{2}  \\ \hline
    \end{tabular}
    }
\end{table}

\begin{table}[!tbp]
    \centering
    \caption{Breakdown of soft failures based on symptoms}
    \label{tab:soft_failures}
    \vspace{-.05in}
    \resizebox{\columnwidth}{!}{
    \begin{tabular}{|c|c|c|c|c|c|}
        \hline
                &  SIGSEGV   & SIGBUS   & SIGABRT   & Other      \\ \hline
        HPCCG   & \num{3322} & \num{32} & \num{22}  & \num{33}   \\ \hline
        CoMD    & \num{2195} & \num{57} & \num{41}  & \num{146}  \\ \hline
        miniMD  & \num{4028} & \num{6}  & \num{25}  & \num{6}    \\ \hline
        GTC-P   & \num{1196} & \num{49} & \num{375} & \num{24}   \\ \hline
    \end{tabular}
    }
    \vspace{-.05in}
\end{table}

\begin{table}[!tbp]
    \centering
    \caption{Latency distribution for soft failures}
    \label{tab:latency}
    \vspace{-.05in}
    \resizebox{\columnwidth}{!}{
    \begin{tabular}{|c|c|c|c|c|c|}
        \hline 
    \multirow{2}{*}{}  & \multicolumn{4}{c|}{Latency (Instructions)} \\ \cline{2-5}
                & $\leq 10$ & $11 \sim 50$ & $51 \sim 400$ &  $> 400$    \\ \hline
        HPCCG   & $99.09\%$ & $0.482\%$  & $0.602\%$   & $0.301\%$   \\ \hline
        CoMD    & $64.15\%$ & $23.57\%$  & $7.43\%$    & $4.85\%$    \\ \hline
        miniMD  & $53.65\%$ & $22.09\%$  & $0.03\%$    & $24.23\%$   \\ \hline
        GTC-P   & $52.68\%$ & $28.76\%$  & $9.7\%$     & $8.86\% $   \\ \hline
    \end{tabular}
    }
    \vspace{-.1in}
\end{table}

We categorized the general outcomes of injections into 4 groups: Benign, Crashes, 
SDC, and Hang. A transient fault is benign (or in short vanishes without causing 
any change in execution) if it doesn't have impact on the application. In such 
cases, the faulty value could either refer to an incorrect but valid memory location 
containing the same value to the original memory location, or its effect is masked 
by a program operation (e.g., min/max operator that masks injections, 
or bit-wise logical operation that suppresses most or least significant bits). Otherwise, 
it will either kill a process (Crash), lead to incorrect outputs (SDC), or result in a 
hang state where there is no progress on execution. As presented in Table~\ref{tab:outcomes}, 
even though majority of faults are benign, around $28.89\%$ of them manifest as crashes, 
and $27.63\%$ of them lead to SDCs. While faults happening in FPU are more 
likely to cause SDCs, the faults manifested in ALU instructions are more likely to 
lead to crashes. Once an application crashes, it 
needs to be restarted incurring costly recovery operations using check-pointed values. 

Table~\ref{tab:soft_failures} breakdowns the crashes based on symptoms. It shows that, 
most ($72.75\% \sim 99.08\%$, $89.8\%$ on average) of crashes manifest as {\it SIGSEGV}, 
typically because they corrupt address calculations and lead applications to access 
invalid memory locations. In addition, Table~\ref{tab:latency} presents the distribution 
of their latency, measured as the number of instructions executed from the fault injection 
point to the crash point. As it shows, the vast majority of crashes ($> 83\%$) were manifest 
within 50 or less dynamic instructions, with more than half of them manifesting within 10 
dynamic instructions. We believe such low-latency manifestation implies that the original 
values (stored in registers or memory) which were involved in the address computation were 
likely to be intact during this latency window, and that it might be possible to recover 
the calculation and essentially mask the fault by creating mechanisms to access these 
original values to recompute the effective address destroyed by the fault.

\subsection{Overall Performance}
In this subsection, we evaluate the performance of \shortnm. In 
general, we are interested in three
questions: 1) how many crashes can \shortnm recover from (recovery 
rate)?; 2) How 
quickly can it recover from a crash?; and 3) What is its overhead during the normal (no fault) run of applications?

\subsubsection{Failure Recovery Rate} 
\begin{figure}
    \centering
    \includegraphics[width=\columnwidth]{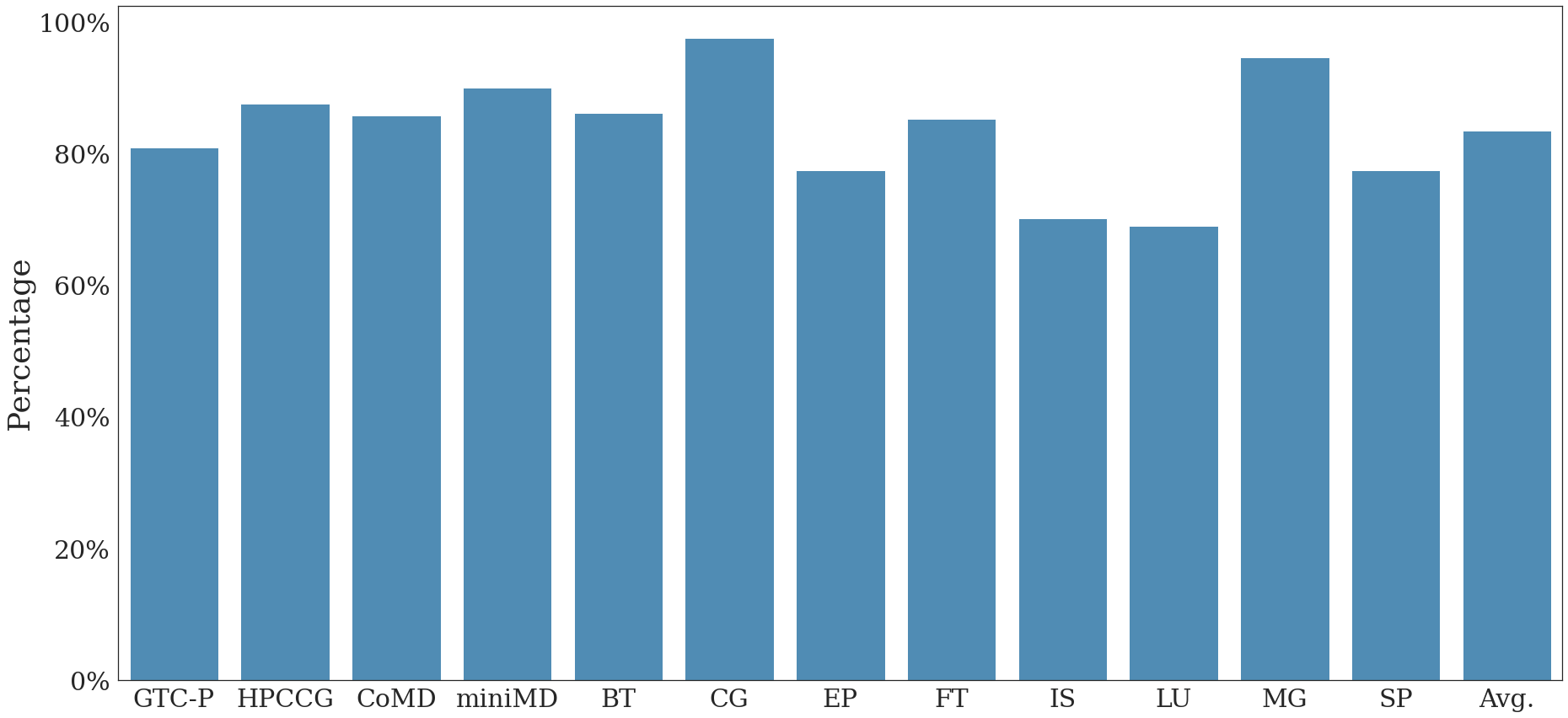}
    \caption{Failure Recovery Rates of \shortnm.}
    % \vspace{-.1in}
    \label{fig:coverage}
\end{figure}

Fig.~\ref{fig:coverage} presents the fault coverage of \shortnm. On average, 
\shortnm can recover $83.55\%$ of injected {\it SIGSEGV} faults, with up to 
$97.6\%$ for CG. \shortnm achieved 
such high fault coverage mainly due to the fact that majority of {\it SIGSEGV} faults manifest quickly, typically 
within only a few dynamic instructions after they occur. The values used in address computations are 
less likely to get updated during such a short time window, especially in the evaluated workloads where 
they are infrequently updated at the algorithm level. Therefore, \shortnm has a good chance to recompute 
the addresses. It is worth noting that during a recovery of failure, \shortnm will not substitute silent 
data corruptions (SDCs) for failures as is possible with more  heuristic based recovery methods~\cite{FaSM2014}. 
This is because the computation of a recovery kernel is based on the raw data fetched from the process.
If raw data is contaminated by a fault, the recovery kernel will definitely generate a wrong address
which is the same as the one accessed by the corrupted instruction. Otherwise, \shortnm is guaranteed 
to get correct address, since it exactly clones the original address computation from applications.

\subsubsection{Recovery Time}
\begin{figure}
    \centering
    \includegraphics[width=\columnwidth]{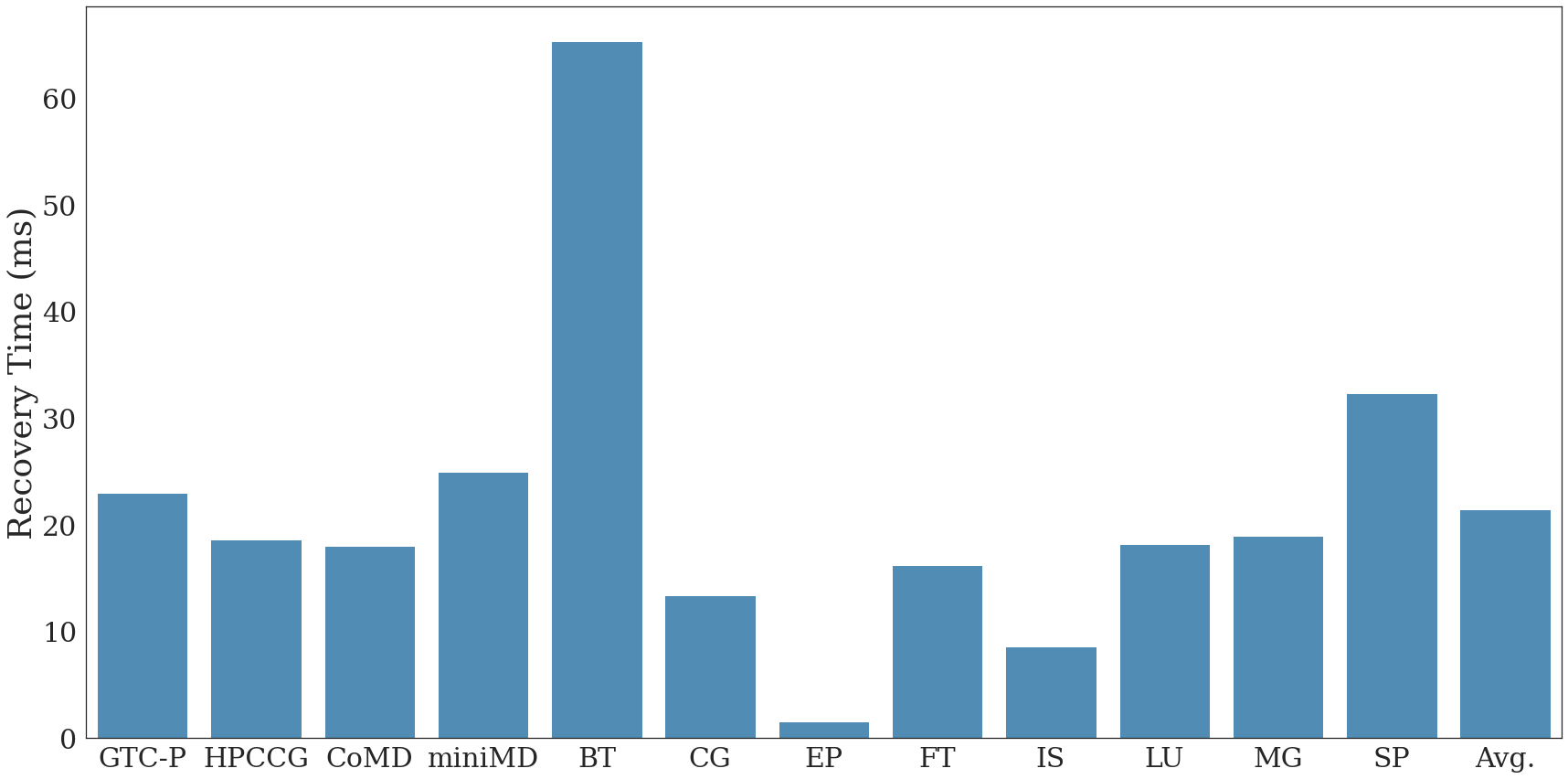}
    \caption{Recovery time of \shortnm}
    \vspace{-.1in}
    \label{fig:latency}
\end{figure}

Recovery time measures the time required by \shortnm to recover from a crash. Clearly a single faulted 
computation might feed into several memory access instructions. What might not be intuitively obvious 
is that in this situation, the \safeguard could be activated several times, recovering the effects of 
each manifestation of the fault. Fig.~\ref{fig:latency} shows that \shortnm can recover a process 
from a {\it SIGSEGV} fault with only a few tens of milliseconds. In fact, only a tiny percentage of 
that recovery time is spent in the generated recovery kernel. They generally only contain a few 
instructions related to address computations and while their use is key to \shortnm, their actual 
portion of the recovery time is negligible. For each activation, more than $98\%$ of the recovery 
time is spent on preparing the execution of recovery kernels, including diagnosing the failure, 
loading recovery table and recovery library, and retrieving arguments from stalled process. 

\subsubsection{Runtime Overheads}
\begin{figure}
    \centering
    \includegraphics[width=\columnwidth]{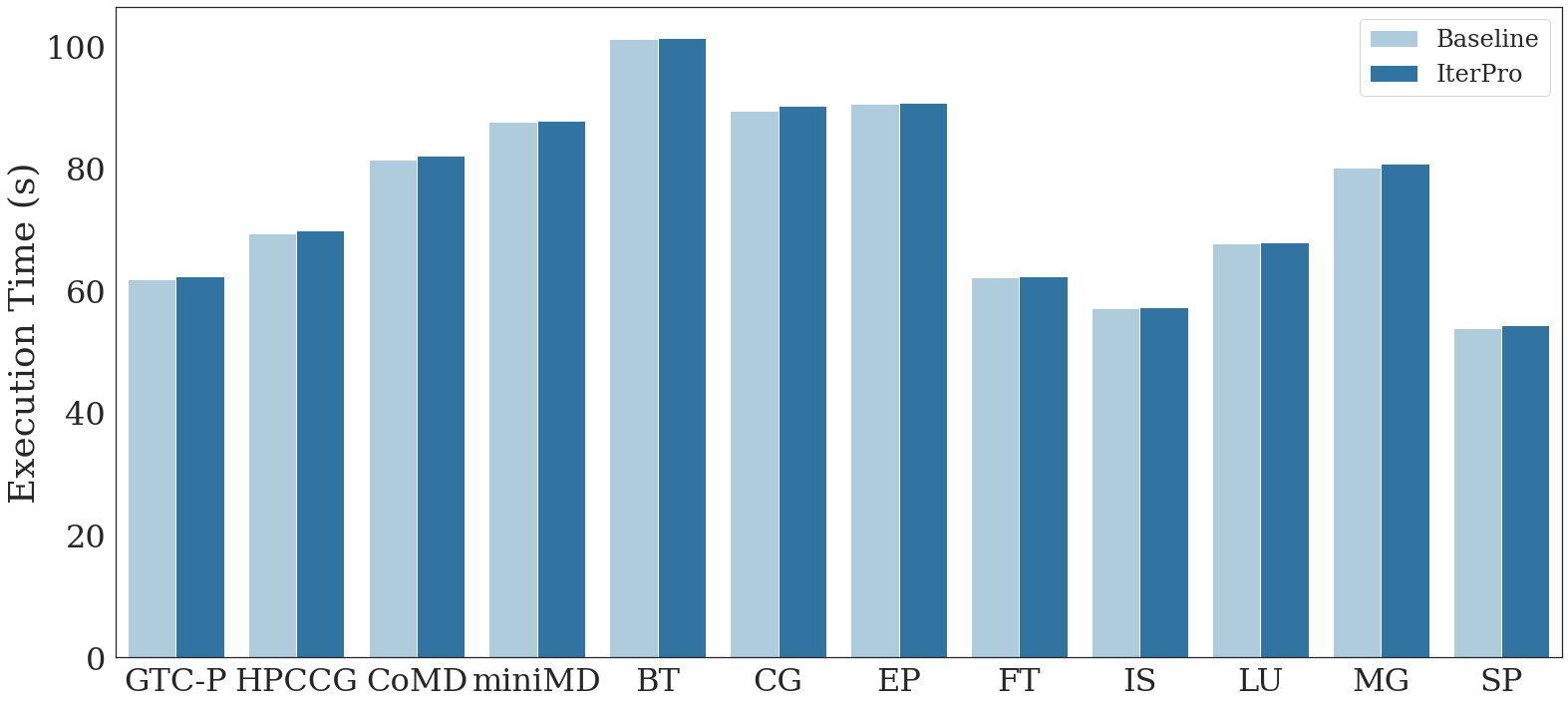}
    \vspace{-.2in}
    \caption{Runtime overhead of \shortnm}
    % \vspace{-.1in}
    \label{fig:overhead}
\end{figure}

\shortnm runtime system and recovery kernels don't reside in normal execution paths of the 
application and are actually loaded dynamically in the case of a fault. Therefore \shortnm's recovery 
mechanism has no performance interference on normal runs of applications, except that it consumes 
a fixed size of main memory ($27$MB, $<1\%$ for evaluated workloads). However, the  LLVM passes 
that enhance recover-ability through independent compute promotion and micro-checkpoint potentially may 
have some minor impacts. They could slightly increase register pressure and introduce more memory-to-register 
data movements, therefore impact the binary code performance. However, these effects are likely to be 
negligible or non-existent depending upon exact details of code and architecture. Fig.~\ref{fig:overhead} 
compares execution times for binaries compiled from baseline (code compiled with classic compilation flag) 
and \shortnm transformed codes. It shows these two set of binaries almost have the same execution times 
(with around $0.51\%$ differences), which implies that these effects are too small to be easily detectable 
on whole application runs with any of our example applications. Similarly, parallel execution times are also unaffected under \shortnm.

\subsection{Efficiency of Novel Code Transformations}
In this subsection, we evaluate the utility of the introduced code transformations by comparing the recovery 
rate of two different setups: 1) a baseline evaluation of CARE when induction variables are not protected (these results correspond to our SC2019 paper~\cite{CGSQ2019}); and 2) a comprehensive evaluation when the code transformations are applied, and induction variables are protected. They are respectively labeled as \textbf{CARE} and \shortnm. In this experiment, 
results for GTC-P, HPCCG, and NPB benchmark suites (exclude CG) are presented since these extensions bring almost no 
improvements for CoMD, miniMD and CG.

\begin{figure}
    \centering
    \includegraphics[width=\columnwidth]{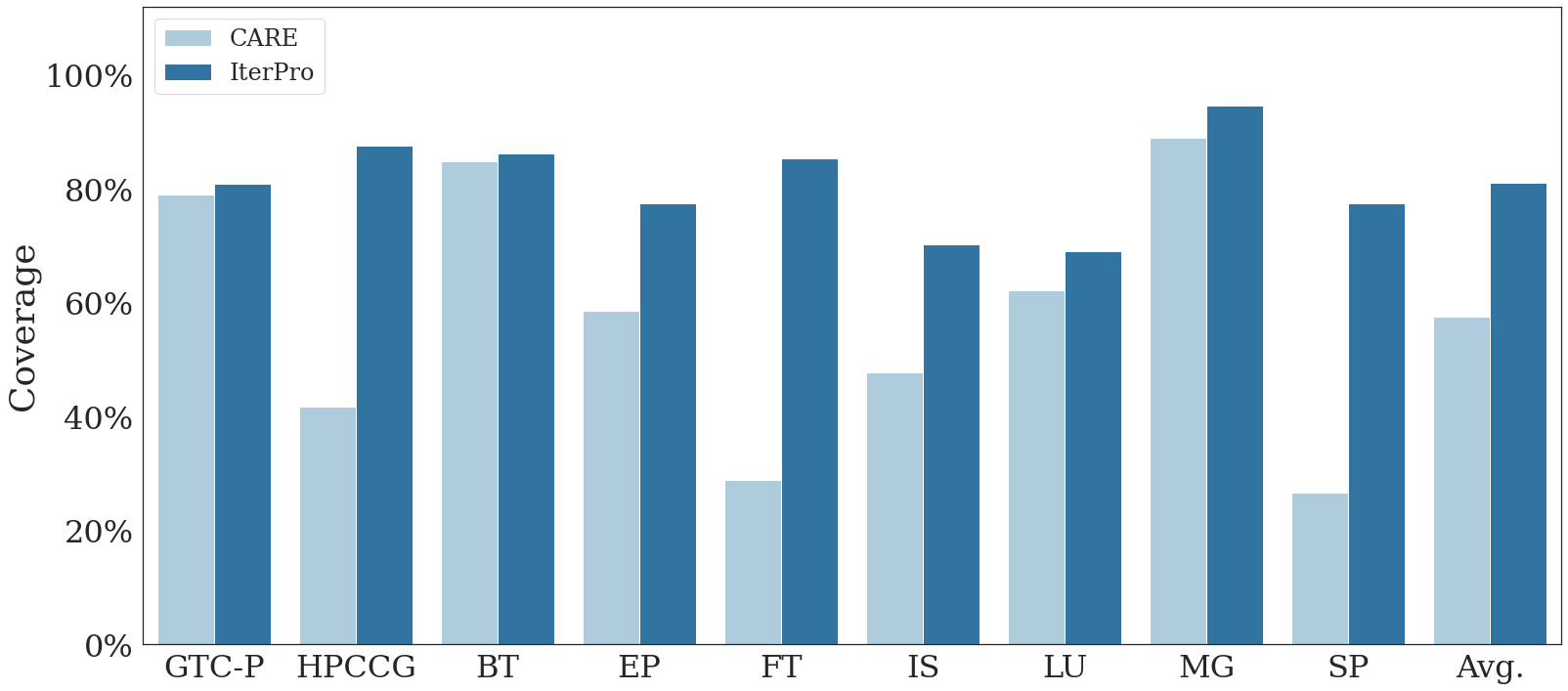}
    \vspace{-.2in}
    \caption{Failure Recovery Rates if different schemes. It shows the advantage 
    of exploiting side-effects of code optimizations and the efficiency of \shortnm 
    code transformations.}
    \vspace{-.1in}
    \label{fig:ct}
\end{figure}

Fig.~\ref{fig:ct} presents the failure recovery rate for each considered scheme. As shown in the figure, \shortnm 
improved recovery rate for $9$ out of $12$ evaluated benchmarks. For these 9 workloads, \shortnm can  recover $81\%$ of 
injected {\it SIGSEGV} faults on an average, while the CARE only recovers $57.64\%$ of these failures. For $3$ of them, including 
FT, SP and HPCCG, \shortnm improved the recovery rate by more than $2\times$. On an average, it improved recovery rate by $1.6\times$ 
across these 9 benchmarks. \shortnm can achieve such significant improvements mainly because of its ability to recover from 
corruptions in induction variables, which is not available 
in CARE. It shows that the proposed extensions discussed in section 3.2 to the normal 
LLVM code generation are key to the success of \shortnm in that they 
significantly add to the set of faults from which \shortnm can recover by introducing ``partner'' induction variables for 
some cases where none naturally exist, and storing away necessary initial values where they would not have been otherwise 
available. In other words, they introduce more ``recoverable''  variables into codes increasing their resilience. 
Table ~\ref{tab:statistics} shows the impact of introduced code transformations by comparing the number of recoverable induction 
variables in original codes and in \shortnm generated codes. As shown in the table, \shortnm's additional LLVM passes 
increased the number of recoverable induction variables by $4\% \sim 500\%$ ($72.65\%$ on average) for $7$ workloads. For 
two others, including EP and IS from NPB, \shortnm's additional code transformations introduce a recovery opportunity for 
induction variables where none existed before (marked by BIG).

\begin{table}
    \centering
    \caption{Number of recoverable induction variables respectively in 
             original and \shortnm transformed codes.}
    \label{tab:statistics}
    \begin{tabular}{|c|c|c|c|c|}
        \hline
         Benchmark  &  \# of Loops  &   Original   &  \shortnm  & Improvement \\ \hline
          GTC-P     &      333      &      145     &     167    &  $15.17\%$  \\ \hline
          HPCCG     &       30      &       38     &      43    &  $13.16\%$  \\ \hline
            BT      &      177      &      253     &     277    &  $9.49\%$   \\ \hline
            CG      &       38      &       8      &      40    &  $500\%$    \\ \hline
            EP      &       12      &       0      &       4    &  $BIG$        \\ \hline
            FT      &       53      &       46     &      48    &  $4.35\%$   \\ \hline
            IS      &        7      &       0      &      12    &  $BIG$        \\ \hline
            LU      &      189      &      340     &     370    &  $8.82\%$   \\ \hline
            MG      &       81      &       32     &      64    &  $200\%$    \\ \hline
            SP      &      316      &      364     &     474    &  $30.22\%$  \\ \hline
    \end{tabular}
\end{table}

\section{Related Work}\label{sec:relate}
Recovery from failures is getting increasing attention in HPC and other 
environments exist~\cite{Chen2011, Chen2013, FGDP2017, CSOG2017, CGMS2018}. 
In this section, we present a brief survey of prior work that is most related to~\shortnm.

% Studies in~\cite{LiVY2012,CGBM2014,CSOG2017} examined the impact of transient faults on 
% scientific applications. Their results showed that a significant portion of transient 
% faults could  manifest as soft failures. This motivated us to study how soft failures 
% manifest inside scientific applications and whether there are common features that can 
% be explored to design an efficient resilience mechanism for them, resulting in the design 
% of \shortnm.~\citet{GLNS2017} and~\citet{CLKS2018} designed and evaluated new fault 
% injection tools for transient faults. While these tools are valuable to the community, 
% they are not a good fit for \shortnm because they either work on high-level intermediate 
% representations (LLVM machine IR) which is inaccurate as compared binary-level injections, 
% or incur high overheads (3$\times$ slowdown) making it infeasible to run large scale 
% ($\sim$\num{100000} injections) fault injection experiments.

There are several studies on online recovery from process failures 
such that applications can continue their normal executions. Rx~\citep{QTSZ2005} aims to recover 
from a process failure by rolling applications back to a previous safe status, and then 
continuing its execution with a minor modification to its environment. Rx is motivated by 
the observation that many program bugs are associated with the setup of process environments, 
so changing the environment setup could avoid the crashes. Its techniques could help handle 
transient faults by simply replaying the computation {\it without} changing the environment, 
however its basic operation requires at least partial application checkpoints which are likely 
to have significant cost. 
RCV~\citep{FaSM2014} is another online failure recovery technique for divide-by-zero (\textit{SIGFPE}) 
and null-dereference (\textit{SIGSEGV}) errors. RCV's approach explores a set of heuristics for recovery. 
For instance, it returns zero as the default result of the divide for divide-by-zero errors, discards 
invalid write instructions that accessing near-to-zero addresses and returns 0 for invalid read operations. 
These techniques are computationally inexpensive and may succeed in getting the application to continue, 
but are likely to introduce SDCs as a side effect. LetGo~\citep{FGDP2017} shares a similar idea to RCV, 
and is specially designed for handling soft failures in scientific applications. Its recovery strategy 
employs a set of heuristics too. Upon a failure, it will reset architecture states to a pre-defined 
value, and then continue the execution of the application. Obviously such heuristic based method could 
lead to SDCs which could be hugely problematic due to incorrect outputs. 

In contrast, \shortnm undertakes a proper recovery process with regards to the maligned address computation 
by recomputing it as per the program semantics and through the use of un-tainted values by synthesizing a very 
lightweight function. It develops careful correspondence mechanism to co-relate the recovery handlers to the 
fault causing instruction at runtime. While \shortnm shares the similar goal and design to RCV and LetGo in that 
they all aim to help applications to survive failures by replacing the default signal handler with their own 
to provide recovery services, \shortnm's approach is superior to others, and will not introduce 
SDCs. \shortnm extends CARE~\cite{CGSQ2019} with the capability of 
recovering crashes due to the corruption in induction variables by exploiting the side effects introduced by modern compiler 
optimizations leading to a significant increase in fault coverage.
\section{Conclusion and Future Work}\label{sec:conclusion}
%Resilience is projected to be a critical challenge for HPC systems due to system scaling trends 
%in higher circuit density, smaller transistor size and near-threshold voltage(NTV) operations. 
%These technology trends would make the system more susceptible to transient errors caused by, 
%e.g., high-energy particle strikes and heat flux. 
Transient errors could not only lead scientific 
applications to generate incorrect outputs, but also crash the execution of an application which 
requires the application to be restarted from the latest checkpoint, and to redo the lost computation. Such approaches could suffer
from high overheads under no-fault execution conditions and could also lead to high downtime required to restore the state. 
In this paper, we present and evaluate \shortnm, a lightweight and compiler-assisted recovery technique 
that allows processes to survive crashes caused by certain transient errors, such that the applications 
can continue their execution. \shortnm is motivated by our observation that {\it SIGSEGV} is a major 
outcome of the transient-error-induced crashes. Thus, for each memory access instruction that involves 
complex address computations, \shortnm will build a recovery kernel by cloning its address
computations. At runtime, it maps the fault causing instruction to a failure recovery handler
which recomputes the address and masks the fault. \shortnm exploits semi-redundancies introduced 
by modern compiler optimization techniques to improve its performance by proposing two new 
code transformations. We evaluated \shortnm with four scientific workloads and 8 benchmarks from the NPB benchmark suites. 
During their normal executions, \shortnm incurs almost {\bf zero} runtime overhead and fixed 27MB memory 
overheads. On an average, \shortnm can recover  $83\%$ {\it SIGSEGV} faults within a few milliseconds, which is a significant improvement as compared to CARE's $57.64\%$ recovery rate.

% use section* for acknowledgment
% \section*{Acknowledgment}
% The authors would like to thank...

% Can use something like this to put references on a page
% by themselves when using endfloat and the captionsoff option.
\ifCLASSOPTIONcaptionsoff
  \newpage
\fi

\bibliographystyle{IEEEtranN}
\bibliography{IEEEabrv,refs}

% biography section
% 
% If you have an EPS/PDF photo (graphicx package needed) extra braces are
% needed around the contents of the optional argument to biography to prevent
% the LaTeX parser from getting confused when it sees the complicated
% \includegraphics command within an optional argument. (You could create
% your own custom macro containing the \includegraphics command to make things
% simpler here.)
%\begin{IEEEbiography}[{\includegraphics[width=1in,height=1.25in,clip,keepaspectratio]{mshell}}]{Michael Shell}
% or if you just want to reserve a space for a photo:

\begin{IEEEbiography}[{\includegraphics[width=1in,height=1.25in,clip,keepaspectratio]{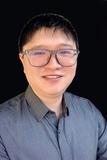}}]{Chao Chen}
is a Software Engineer at Amazon. He got his Ph.D. 
from Georgia Institute of Technology under the supervision of Greg Eisenhauer and Santosh Pande. 
He interested in building systems and compiler techniques. His thesis works on lightweight 
resilience mechanisms for extreme-scale HPC 
systems by exploring program features via 
compiler techniques.
\end{IEEEbiography}
%
% if you will not have a photo at all:
\begin{IEEEbiography}[{\includegraphics[width=1in,height=1.25in,clip,keepaspectratio]{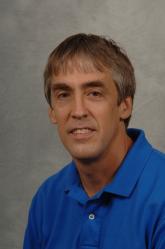}}]{Greg Eisenhauer}
is a senior research scientist in the College of Computing at the Georgia Institute of Technology and Technical Director of the Center for Experimental Research in Computer Systems. He received the B.S. degree in Computer Science (1983) and the M.S. degree in Computer Science (1985) from the University of Illinois, Urbana-Champaign. He received the Ph.D. degree from the Georgia Institute of Technology in 1998. His research is about HPC systems. 
\end{IEEEbiography}
%
% % insert where needed to balance the two columns on the last page with
% % biographies
% %\newpage
\begin{IEEEbiography}[{\includegraphics[width=1in,height=1.25in,clip,keepaspectratio]{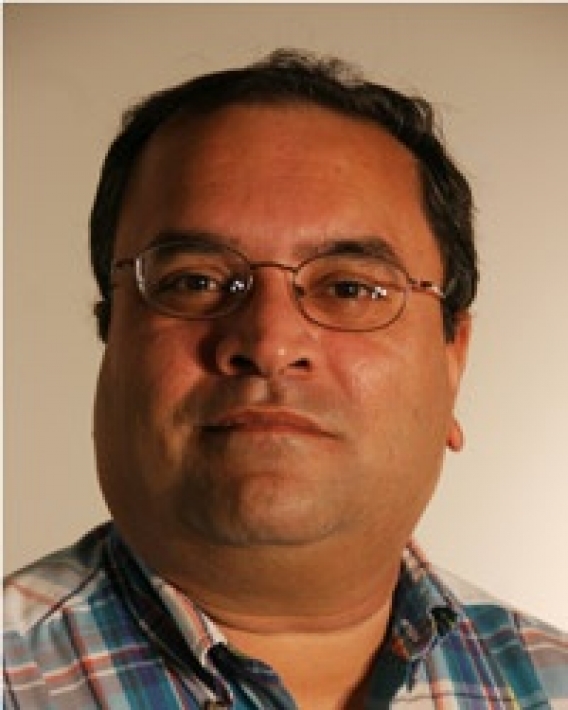}}]{Santosh Pande}
is a Professor in the School of Computer Science, College of Computing at the Georgia Institute of Technology. Pande's primary interest is in investigating static and dynamic compiler optimizations on evolving architectures. His research philosophy involves tackling practical problems which are relevant and important to the current issues in systems research and propose foundational solutions to them for good impact.
\end{IEEEbiography}

% You can push biographies down or up by placing
% a \vfill before or after them. The appropriate
% use of \vfill depends on what kind of text is
% on the last page and whether or not the columns
% are being equalized.

%\vfill

% Can be used to pull up biographies so that the bottom of the last one
% is flush with the other column.
%\enlargethispage{-5in}

% that's all folks
\end{document}